# Efficient electrical switching of exciton states for valley contrast manipulation in two-dimensional perovskite/monolayer $WS_2$ heterostructures

Yingying Chen[1], Zeyi Liu[1], Junchao Hu[1], Junze Li[1], Wendian Yao[1], Dehui Li[1,2*]

[1]School of Optical and Electronic Information, Huazhong University of Science and Technology, Wuhan, 430074, China

[2]Wuhan National Laboratory for Optoelectronics, Huazhong University of Science and Technology, Wuhan, 430074, China

*Correspondence to: Email: dehuili@hust.edu.cn.

**Abstract**

The coupled spin-valley physics in transition metal dichalcogenides (TMDs) endows exciton states with valley degrees of freedom, making them promising for valleytronic applications in TMDs monolayers and/or their heterostructures. Although the valley dynamics of intralayer and interlayer excitons (IXs) have been studied, efficient manipulation of valley pseudospins by switching exciton states remains elusive. Therefore, it is of great importance to effectively tune the exciton states to obtain practical valley polarization switches for valley encoding. Here, we demonstrate the electrical switching of exciton emission with highly variable valley polarization mediated by charged IXs (CIXs) in the heterostructure of monolayer $WS_2$ and two-dimensional (2D) perovskite, irrespective of lattice constants, the rotational and translational alignment. The formation of IXs is identified by photoluminescence excitation (PLE) and photoluminescence (PL) studies, which can be further electrically tuned from positively charged to negatively charged depending on the electrostatic doping level of monolayer $WS_2$. Importantly, we demonstrate an electrical switching from type-II to type-I band alignment, manifesting as a change in the PL profile from CIX to charged intralayer exciton emission. Such transition induces a large contrast in valley polarization between the two exciton states, enabling the reversible electrically regulated valley polarization switch with a maximum ON/OFF ratio of 15.8. Our study provides an alternative mechanism to achieve valley polarization switching with great simplicity for valleytronics and the electrical control of exciton species and associated valley-contrasting physics would further facilitate the development of optoelectronic and valleytronic devices.

**Introduction**

Monolayer transition metal dichalcogenides (TMDs) have emerged as promising candidates for valleytronic devices due to the spin-valley locking effect, which introduces two inequivalent K and K′ valleys at the corner of the hexagonal Brillouin zone with opposite valley pseudospins[1-4]. Stacks of TMDs provide a platform for realizing different exciton states with highly tunable valley dynamics. Interlayer excitons (IXs) in TMD heterobilayers[5], formed by electrons and holes residing in separated layers in type-II band heterostructures[6,7], are reported to have an optically addressable valley degree of freedom[8,9]. Compared with intralayer excitons, IXs exhibit a longer depolarization lifetime[10], leading to a larger degree of valley polarization[11], which is essential for valleytronic devices. Although previous works have demonstrated that the valley polarization of intralayer excitons in monolayer TMDs can be manipulated by electrostatic doping[12,13]; nevertheless, inconspicuous variations in valley contrast make it difficult to manipulate the valley encoding to achieve information input and readout for valleytronic applications[14,15]. In contrast, the valley dynamics of IXs show a strong gate dependence which introduces an intense asymmetry in valley polarization by changing gate voltages[16-18]. Moreover, considering that singlet and triplet IX emission are equipped with opposite spin configurations in a moiré superlattice[19], an electrically tunable valley polarization switch based on IXs has been demonstrated[20]. However, such observation yields TMD heterobilayers with angles near 0°or 60°to brighten the forbidden optical transitions[21,22], thus increasing costs and limiting the development of valleytronics based on TMD heterobilayers.

A general strategy is to explore new layered materials stacking with TMDs to achieve direct interlayer transitions regardless of lattice constants, stacking alignment and whether IXs are momentum direct or indirect[23-26]. Ubrig et al. have demonstrated broad-spectrum responses based on IXs in $n$L-InSe/2L-WS$_2$ heterostructures where the interlayer transitions locate at the Γ point involving $n$L-InSe conduction bands and 2L-WS$_2$ valence bands[24]. Tan et al. have reported direct IX emission in multilayer WSe$_2$/MoS$_2$ heterostructures with up to four layers of MoS$_2$[25]. Our previous work has also observed robust IX emission in various two-dimensional (2D) perovskite/monolayer $WSe_2$ heterostructures irrespective of stacking angles and momentum mismatch[26]. Nevertheless, valley information in those heterostructures either loses or requires to stack with chiral 2D perovskites to achieve valley manipulation[27]. To this end, it is essential to explore a new mechanism to manipulate the valley polarization of IXs in non-TMD/TMD heterostructures.

Here, we demonstrate the electrically tunable valley polarization by switching the exciton states from charged IXs (CIXs) to charged intralayer excitons. The presence of IXs in the monolayer $WS_2$/2D perovskite heterostructure is validated by photoluminescence excitation (PLE) and power- and temperature-dependent photoluminescence (PL) studies. An electrical manipulation could efficiently switch between the charged intralayer excitons of $WS_2$ and CIXs of heterostructure due to a prominent switch from type-II to type-I band alignment, which exhibits a strong asymmetry in valley contrast, allowing us to achieve electrically controllable valley polarization switching for logic operations.

**Results and Discussion**

Figure 1a illustrates the schematic of the $WS_2$/(iso-BA)$_2$PbI$_4$ heterostructure. The (iso-BA)$_2$PbI$_4$ single crystals (iso-BA is iso-butylamine with a chemical formula of $CH_3(CH_2)_3NH_3$) were synthesized by solution methods reported in our previous work[28]. Thin (iso-BA)$_2$PbI$_4$ flakes and $WS_2$ monolayers were prepared by mechanical exfoliation from their bulky crystals. The heterostructure was fabricated by dry-transfer method onto $SiO_2$ (300 nm)/Si substrate where the monolayer $WS_2$ was placed on the top of (iso-BA)$_2$PbI$_4$ flake. According to previous studies[29,30], the energy band between (iso-BA)$_2$PbI$_4$ and monolayer $WS_2$ is type-II alignment (Fig. 1b). Figure 1c shows the optical image of the as-prepared heterostructure and corresponding fluorescence images with different long-pass filters at room temperature. In the heterostructure region, while a slight PL enhancement of (iso-BA)$_2$PbI$_4$ is observed under a 520 nm long-pass filter (middle panel of Fig. 1c), an obvious PL quenching of monolayer $WS_2$ can be seen with a 590 nm long-pass filter (bottom panel of Fig. 1c). Whereas the enhanced PL of (iso-BA)$_2$PbI$_4$ can be ascribed to the surface protection of $WS_2$ monolayer against unstable 2D perovskite in the ambient condition (Supplementary Fig. 1), the significantly reduced PL of monolayer $WS_2$ suggests efficient charge transfer owing to the staggered band alignment in the $WS_2$/(iso-BA)$_2$PbI$_4$ heterostructure[31,32].

Figure 1d depicts the PL spectra of the $WS_2$/(iso-BA)$_2$PbI$_4$ heterostructure and individual constituent layers at 78 K. The enhanced free exciton emission of (iso-BA)$_2$PbI$_4$ (label as 'X')[28] under a 473 nm (2.62 eV) laser excitation as well as completed quenched neutral exciton and negative trion of WS$_2$ (label as 'X$_0$' and 'X$^-$')[33] under a 532 nm (2.33 eV) laser excitation in the heterostructure region are highly consistent with the observation in Fig. 1c. Especially, a remarkable broad emission (label as 'IX') centered at 1.89 eV with a full width at half maximum of ~130 meV shows a stronger intensity than the intralyer exciton emission of WS$_2$ that also appears in the heterostructure. We expect that this broad emission originates from IXs according to relative band alignment where electrons and holes tend to transfer to the lower energy band to form spatially IXs upon laser exposure (orange oval in Fig. 1b). Figure 1e further compares the PL mapping of individual WS$_2$ and the heterostructure under a 532 nm laser excitation in the cyan box of Fig. 1c, in which the uniform intense IX emission indicates efficient charge transfer and robust IX formation in the WS$_2$/(iso-BA)$_2$PbI$_4$ heterostructure.

**Evidence of charge transfer to form IXs**

Unlike Zhang et al. previously reported in the WS$_2$/(PEA)$_2$PbI$_4$ heterostructure where the PL of monolayer WS$_2$ shows great enhancement as it interfaced with (PEA)$_2$PbI$_4$ due to the energy transfer from (PEA)$_2$PbI$_4$ to WS$_2$[34], the PL of WS$_2$ is completely quenched in our experiments while the IX emission is rather prominent, indicating the occurrence of charge transfer in a type-II heterostructure. In addition, the charge transfer process can be distinguished from energy transfer by the insertion of an insulating hexagonal born nitride (*h*-BN) layer[6]. When a 6 nm-thickness *h*-BN layer is inserted into the WS$_2$/(iso-BA)$_2$PbI$_4$ heterostructure where the charge transfer is suppressed but energy transfer can still take place[35], we observe the disappeared IX emission in the WS$_2$/*h*-BN/(iso-BA)$_2$PbI$_4$ heterostructure region, suggesting that the *h*-BN layer hinders charge transfer and thus blocks the formation of IXs (Supplementary Fig. 2). This observation suggests that the energy transfer is not responsible for this broad emission in the WS$_2$/(iso-BA)$_2$PbI$_4$ heterostructure, otherwise, it should persist in the WS$_2$/*h*-BN/(iso-BA)$_2$PbI$_4$ heterostructure. Instead, charge transfer dominates the WS$_2$/(iso-BA)$_2$PbI$_4$ heterostructure, forcing electrons and holes to transfer to separated layers to form IXs. Moreover, the energy transfer can also be excluded *via* the PL study under the excitation of the 532 nm laser (Supplementary Section 2).

Furthermore, we have performed PLE measurements to strengthen the formation of IXs involving two constituent layers. The PL spectra of IX emission are recorded as a function of excitation laser energy (Fig. 2a), which shows multiple peaks when the excitation is in resonance with different exciton absorption features of $WS_2$ and $(iso-BA)_2PbI_4$. To assign the peaks of the PLE spectrum, we have measured the absorption spectra of $WS_2/(iso-BA)_2PbI_4$ heterostructure and the individual constituent layers on the quartz substrate in Fig. 2b. While the $(iso-BA)_2PbI_4$ exhibits a ground X absorption peak at ~2.37 eV[36], monolayer $WS_2$ shows three typical absorption peaks corresponding to A, B, and C excitons[37], respectively. In the heterostructure region, the A exciton peak of $WS_2$ is redshifted by ~44 meV and the X exciton peak of $(iso-BA)_2PbI_4$ is slightly blueshifted by ~6 meV due to the robust interlayer coupling[38], and the broader linewidths of these excitonic peaks are attributed to a charge transfer effect that introduces nonradiative transitions[39]. By comparing the absorption spectra with the PLE spectrum of heterostructure in Fig. 2b, we can ascribe the PLE peaks at 2.04 eV, 2.42 eV, and 2,88 eV to the A exciton of $WS_2$, X exciton of $(iso-BA)_2PbI_4$ and C exciton of $WS_2$, respectively[34]. To this end, the PLE spectrum of the $WS_2/(iso-BA)_2PbI_4$ heterostructure resonates with excitonic transitions of both $WS_2$ and $(iso-BA)_2PbI_4$ indicating the formation of IXs involving charge transfer between the two layers (Supplementary Fig. 3).

**Optical characterizations of IXs**

We have also performed power- and temperature-dependent PL studies to characterize interlayer coupling in the $WS_2/(iso-BA)_2PbI_4$ heterostructure. Figure 2c shows that as the incident power increases, the peak energy of IX emission gradually blueshifts and finally saturates, which can be interpreted as the repulsive dipole-dipole interaction between oriented IXs[7,40]. By fitting the PL intensity $(I)$ as a function of excitation power $(P)$ through a power-law relation, $I \propto P^\alpha$, IX emission intensity versus excitation power exhibits a sublinear increase with a slope of 0.7, which agrees well with previous reports[7,26] (Fig. 2d). Figure 2e shows that IX emission gradually redshifts as temperature increases and disappears above 240 K, along with redshifted intralayer exciton emission of $WS_2$ and reduced $X^-/X_0$ ratio at high temperature (Supplementary Fig. 5), indicating the typical bandgap reduction with increasing temperature in semiconductors[26,33] (Supplementary Section 3). All these results are consistent with IXs observed in our previously reported $(iso-BA)_2PbI_4/WSe_2$ heterostructures[26], suggesting the common formation of IXs in heterostructures of monolayer TMDs and 2D perovskites.

It's worth noting that the existence of IXs is independent of the stacking sequence of (iso-BA)$_2$PbI$_4$ and WS$_2$. We have compared two opposite constructions of (iso-BA)$_2$PbI$_4$ and WS$_2$ in Supplementary Fig. 6, in both of which IX emission blueshift with increasing incident power and exhibit similar sublinear slopes. The blueshifted peak energy and weaker PL intensity of IXs in the heterostructure with WS$_2$ monolayer on the bottom can be explained as a changing dielectric environment[41] and substrate-induced-doping of monolayer WS$_2$[42]. Moreover, we also observe that IXs are not affected by excitation energy, which shows similar IX emission profiles and exhibits similar sublinear slopes around 0.7 under different laser excitation (Supplementary Fig.7). In addition, by altering the organic cations of 2D perovskites to change the interlayer distance, IXs show similar PL profiles but with different PL intensity, peak energy, and sublinear slopes due to different coupling strengths (Supplementary Fig. 8). We also exclude the possibility of defect states contributing to the broad emission in the heterostructure by comparison with aged monolayer WS$_2$, WS$_2$/(iso-BA)I hybrid and the type-I WS$_2$/(BA)$_2$PbBr$_4$ heterostructure (Supplementary Section 5 and Supplementary Figs. 9-11). Therefore, our results validate the formation of IXs in the heterostructure of various monolayer TMDs and 2D perovskites.

Figure 2f further shows the lifetime of IXs in the WS$_2$/(iso-BA)$_2$PbI$_4$ heterostructure. A fit to a single exponential decay yields the IX lifetime of 1.96 ns which is comparable with the previous report[7]. As we discussed before, to form the lowest energy configuration of IXs, electrons and holes are spatially separated, thus reducing the overlap of the wavefunction of electrons and holes and leading to the long lifetime of IXs[43]. The nanosecond-lifetime scale of IXs is two orders of magnitude larger than the lifetime of intralayer excitons in monolayer TMDs[44], further facilitating the application of IXs in optoelectronic devices.

**Electrical control of IX emission**

To explore the optoelectronic applications of IXs, we have constructed an $h$-BN-encapsulated (iso-BA)$_2$PbI$_4$/WS$_2$ device on the Si$_3$N$_4$ (100 nm)/Si substrate as depicted in Fig. 3a. The WS$_2$ layer is directly contacted with the few-layer graphene (FLG) sheet for electrostatic doping by applying the gate voltage ($V_g$) between the FLG and highly doped Si substrate and is half covered by the (iso-BA)$_2$PbI$_4$ flake for comparison with the heterostructure region. Figure 3b,c presents the color plot of PL spectra against $V_g$ under a 532 nm laser excitation for WS$_2$ and heterostructure, respectively. The emission band of WS$_2$ shows an evolution from X$_0$ to X$^-$ peak as $V_g$ is swept from –20 V to 20

V. The peak intensity of $X_0$ emission gradually decreases and the peak energy slightly blueshifts, while the $X^-$ emission becomes prominent and redshifts, resulting in an $X^-$ dominated profile at a high electron doping level. These observations can be attributed to phase space blocking and many-body effects according to previous reports on intrinsic n-doped $WS_2$[13,45]. Similarly, the emission band of heterostructure also shows the evolution of various exciton species depending on the doping level of $WS_2$. The IX is initially positively charged at –20 V to form $IX^+$ and then reaches a neutral state around –10 V. Thereafter, IX becomes negatively charged to form $IX^-$ and blueshifts due to the increase in carrier density, analogous to the increase in incident power (Fig. 2c). As the electron doping level further increases, $IX^-$ gradually disappears while $X^-$ dominates the emission at positive $V_g$. Such observation also indicates that IXs formed in undoped heterostructures are naturally negatively charged. Importantly, the stacking order can significantly affect the electrical tunability. When the (iso-BA)$_2$PbI$_4$ is placed beneath the $WS_2$ layer, the heterostructure is difficult to be electrostatically tuned due to the large thickness of 2D perovskite (Supplementary Fig. 12).

To understand the evolution of exciton emission in the heterostructure region, we propose a model of gate-dependent carrier redistribution in Fig. 3d. Noted that charged excitons can be referred to as three-particle trions (bound states of two electrons and one hole, or one electron and two holes). When $WS_2$ is initially p-doped at –20 V, the Fermi energy level ($E_F$) will cross the valance band of (iso-BA)$_2$PbI$_4$ and neutral IX tends to capture an extra hole to form $IX^+$. As $V_g$ increases, the $E_F$ gradually crosses the conduction band of $WS_2$ and becomes n-doped, thus $IX^-$ dominates the heterostructure as Jauregui et al. reported[19]. Further increasing $V_g$ to sufficiently dope $WS_2$, charge accumulation can induce a large energy shift between (iso-BA)$_2$PbI$_4$ and $WS_2$ as Meng et. al demonstrated[46]. In this case, the type-II bandgap alignment will be switched to the type-I band alignment, where both electrons and holes are forced to transfer from the (iso-BA)$_2$PbI$_4$ layer to the $WS_2$ layer, thus facilitating the formation of $X^-$ of $WS_2$ as we observed in Fig. 3c. We have also carried out the PL and PLE studies to support this model. For $V_g$= 20 V, we record the $X^-$ emission for both monolayer $WS_2$ and the type-I heterostructure region. Figure 3e shows the large PL difference upon excitation energy, in which the PL intensity is slightly quenched for the 532 nm laser excitation, but is greatly enhanced for the 473 nm laser excitation in the heterostructure. This is because only the $WS_2$ layer is excited under the 532 nm excitation, while both layers are excited under the 473 nm excitation which allows electrons and holes to transfer from the (iso-BA)$_2$PbI$_4$ layer to the WS$_2$ layer, resulting in the PL enhancement in the heterostructure. In addition, PLE spectra further verify the charge transfer process under different excitation energy. Figure 3f (top panel) presents the PLE spectra for X$^-$ emission for monolayer WS$_2$ and heterostructure at 20 V, respectively. The PLE spectrum of heterostructure clearly shows an additional resonance peak at 2.42 eV compared to monolayer WS$_2$, corresponding to the X exciton absorption of (iso-BA)$_2$PbI$_4$, which suggests charge transfer from (iso-BA)$_2$PbI$_4$ to monolayer WS$_2$, coinciding with the scenario in Fig. 3d. Besides, comparing the PLE spectra of the heterostructure at ±20 V, Figure 3f (bottom panel) shows that the PLE intensity is highly enhanced when the excitation is in resonance with the X exciton absorption of (iso-BA)$_2$PbI$_4$ at 20 V, while the PLE intensity at the B absorption peak of WS$_2$ remains almost unchanged at –20 V. Therefore, the PL enhancement under the 473 nm excitation and the enhanced PLE intensity in resonance with the X exciton absorption of (iso-BA)$_2$PbI$_4$ confirm the type-I band alignment at 20 V, in agreement with the previous work[46]. Such electrical switching of band alignment and associated exciton states favor carrier dynamic studies in TMD and perovskite heterostructures, further facilitating their applications in optoelectronic and valleytronic devices.

We have also explored the electric field control of the heterostructure by placing the FLG sheet on the top of the *h*-BN encapsulated heterostructure as depicted in Fig. 4a, where the electric field can be manipulated by applying a $V_g$ between the top FLG and Si substrate. Figure 4b shows that IX$^-$ remains almost unchanged at negative $V_g$ and gradually disappears along with rising X$^-$ emission at positive $V_g$. We attribute the constant IX$^-$ emission at negative $V_g$ to the dielectric screening effect, probably due to the photoinduced doping effect that screens the electric field modulation[47,48]. The similar PL transition from IX$^-$ to X$^-$ emission at positive $V_g$ as Fig. 3c can be ascribed to the pretty close valance band maximum of WS$_2$ and (iso-BA)$_2$PbI$_4$, where the electric field reduces the band offset between each other when the dipole moment is antiparallel to the direction of the electric field[7]. This assumption is also verified by the device in Supplementary Fig. 13a, where the FLG covers the whole *h*-BN encapsulated heterostructure and simultaneously contacts with monolayer WS$_2$. In this case, both electrostatic gating and electric field modulation are taken into consideration, the PL evolution is quite similar to what we observed in Fig. 3c instead of Fig. 4b, indicating the electric field modulation is screened at negative $V_g$ (Supplementary Fig. 13).

Our observation is quite different from those of TMD heterobilayers, where the electrical field modulation of IX emission yields a linear Stark effect due to the defined out-of-plane electric dipole moment[40]. Indeed, we observe a linear shift of IX peak energy with $V_g$ corresponding to the first-order Stark effect in the previously reported (iso-BA)$_2$PbI$_4$/WSe$_2$ heterostructure, where the peak energy of IX emission redshifts at negative $V_g$ and blueshifts at positive $V_g$ (Supplementary Fig. 14). However, because of the larger valance band offset between (iso-BA)$_2$PbI$_4$ and WSe$_2$, the IX emission cannot be switched to intralayer exciton emission under electric field control.

**Valley polarization switching device**

Our previous work has already shown that IXs in achiral 2D perovskite/WSe$_2$ heterostructures cannot retain the valley polarization information as to what in TMD heterobilayers[27]. Taking advantage of this feature and the fact that exciton species can be electrically controlled in the heterostructure of 2D perovskite and WS$_2$, a large valley polarization contrast between IX$^-$ and X$^-$ can be introduced. The degree of polarization (DOP) is introduced as $\mathrm{DOP} = (I_{\sigma^+} - I_{\sigma^-})/(I_{\sigma^+} + I_{\sigma^-})$ to qualify the circularly polarized light emission, where $I_{\sigma^+}$ and $I_{\sigma^-}$ represent the PL intensity of right-hand and left-hand circularly polarized light, respectively. We calculate the valley polarization in terms of the integrated area of the entire spectrum to eliminate the need for complex optical paths to resolve the luminescent energy. Figure 4c compares the circularly polarized PL of the heterostructure at –20 V, 0 V, and 20 V under the $\sigma^+$ polarized 532 nm laser excitation, respectively. Since IX$^-$ does not possess valley polarization and the DOP of X$^-$ rises slowly as $V_g$ increases (Supplementary Fig. 15), the overall emission presents a small DOP of 1.9 % at –20 V and a large DOP of 20.8 % at 20 V, which can introduce a large difference of 11 times. Figure 4d shows the overall DOP as a function of $V_g$. The trend of DOP is similar to the transfer curve of the WS$_2$ field-effect transistor[49], showing an ON/OFF ratio of 11, which is a sixfold enhancement over the ON/OFF ratio of 1.8 for an individual monolayer WS$_2$ device (Supplementary Fig. 15c). Such significant valley polarization variation allows the WS$_2$/(iso-BA)$_2$PbI$_4$ heterostructures to operate as a valleytronic transistor[50] controlled by electrical means. Furthermore, the controllable DOP can enable a $V_g$ regulated valley polarization switching between IX$^-$ and X$^-$ states, as shown in Fig. 4e, where the reversible valley polarization switching between ±20 V exhibit a sharp transition between two logic states with great stability. Noted that the valley polarization switching device in Fig. 4e exhibits a slightly lower ON/OFF ratio of 8.4, which can be ascribed to the hysteresis effect during cyclic voltage sweeping[51].

In addition, unlike the situation where electrostatic doping of $WS_2$ is inefficient to control IXs in the heterostructure when $WS_2$ is placed on the top of $(BA)_2PbI_4$, changing the stacking order of $WS_2$ and $(BA)_2PbI_4$ can only exchange the dipole orientation of IXs without changing the modulation effect under electric field. We have performed the electric field control of the heterostructure in the opposite stacking order to that of Fig. 4 in Supplementary Fig. 16a, which shows $X^-$ emission at negative $V_g$ and $IX^-$ emission at positive $V_g$ (Supplementary Fig. 16b). The circularly polarized PL spectra vary from voltages (Supplementary Fig. 16c) and the DOP versus $V_g$ presents an ON/OFF ratio of 8.6 (Supplementary Fig. 16d). Such valley contrast also enables reversible valley polarization switching between ±20 V (Supplementary Fig. 16e). Furthermore, the valley contrast can also be achieved by electrostatically doped heterostructure in the scenario in Fig. 3, where a similar DOP trend with an even larger ON/OFF ratio of 15.8 and the stable valley polarization switching device has also been demonstrated (Supplementary Fig. 17).

Apart from Vg-controlled valley polarization switching between different exciton species, the valley polarization can also be tuned by changing the incident power, which also enables a reversible switch. The power-dependent circularly polarized PL spectra in Supplementary Fig. 18a show that $IX^-$ dominates the heterostructure at a low excitation power regime and blueshifts with increasing incident power as we discussed in Fig. 2c. Due to the sublinear slope of $IX^-$ and linear slope of $X^-$ (Supplementary Fig. 18b), the DOP of the heterostructure becomes larger and gradually saturates with increasing power because of the enlarged proportion of $X^-$ in the mixture of $IX^-$ and $X^-$ under a sufficiently high power excitation (Supplementary Fig. 18c). In this case, by selecting the modest incident power to ensure the device stability, a reversible valley polarization switching with an ON/OFF ratio of 6.5 is shown in Supplementary Fig. 18d. Therefore, the valley polarization switching mediated by CIXs can also be achieved by changing the incident power, further enriching the manipulation of CIXs in valleytronic devices.

It should be noted that the ON/OFF ratio of the valley polarization switching here is limited by the optical readout since the measurement error for the circularly polarized light is around 1.5-2.0% in our measurement system. In view of the fact that the exciton emission of $WS_2$ can be completely quenched in 2D perovskite/$WS_2$ heterostructures, the ON/OFF ratio of the valley polarization switching could be infinite in principle. Therefore, we expect that a valleytronic transistor with a giant ON/OFF ratio can be achieved based on our 2D perovskite/WS$_2$ heterostructures via electrical readout, which would be of utmost importance for valleytronics.

**Conclusion**

In summary, we have demonstrated the efficient electrical control of CIX and X$^-$ states to achieve valley polarization switching in the WS$_2$/(iso-BA)$_2$PbI$_4$ heterostructure. The CIX can be electrically tuned from IX$^-$ to IX$^+$ when the monolayer WS$_2$ is electrostatically doped in the heterostructure of WS$_2$ and (iso-BA)$_2$PbI$_4$. More importantly, we have observed an electrically controllable PL transition between CIX and X$^-$ states that is interpreted as a switch from type-II to type-I band alignment. Such variation of exciton states allows the manipulation of valley polarization, where the DOP operates as a function of $V_g$ as a valleytronic transistor with a maximum ON/OFF ratio of 15.8, introducing two logic states between ±20 V, and thus achieving reversible valley polarization switching with great stability. Our results are highly instructive for optoelectronics and valleytronics as they enable the electrical manipulation of different exciton states with large valley polarization contrast for logic operations, and further strengthen our fundamental understanding of IXs in heterostructures of TMDs and 2D perovskites.

**Online content**

Fluorescence images, PL spectra of different samples, power and temperature-dependent PL spectra of different samples, PL intensity fitting of different samples, band alignment of monolayer WS$_2$ and (BA)$_2$PbBr$_4$, gate-dependent PL and PLE spectra of different samples, polarization-resolved PL spectra of different heterostructures under various $V_g$ and incident power, DOP of different samples as a function of $V_g$; details of author contributions and competing interests; and statements of data and code availability are available at ***.

**References**

... 

**Methods**

**Sample preparations:** Au (50 nm)/Cr (10 nm) electrodes were prefabricated onto Si$_3$N$_4$ (100 nm)/Si wafers by photolithography and thermal evaporation. The 2D perovskite bulk crystals were synthesized by solution methods according to the previous work[28]. WS$_2$, *h*-BN, and graphene crystals were purchased from 2D semiconductors. 2D perovskite, monolayer WS$_2$, *h*-BN, and FLG flakes were mechanically exfoliated by the Scotch tapes from their bulk crystals and then transferred onto substrates using PDMS (polydimethylsiloxane) stamps via the dry-transfer method. The thickness of *h*-BN (16 nm for the top layer and 14 nm for the bottom layer) and (iso-BA)$_2$PbI$_4$ (92 nm) in Fig. 4a were measured by atomic force microscopy. All samples were not treated by thermal annealing to maintain the stability of 2D perovskites.

**Optical measurements:** Optical images and fluorescence images were captured on the microscope (Olympus BX53M) with different fluorescence modules (U-FWBS and U-FWGS). Samples were encapsulated in a liquid nitrogen bath cryostat (Cryo Industries of America Inc.) with a $10^{-7}$ Torr vacuum. For PL measurement, The PL spectra were acquired on a home-built Raman spectrometer (Horiba iHR-550) with a 600 g/mm grating at 78 K unless stated otherwise. The excitation energy for every spectrum was as specified in the main text and the laser power was kept at 8.67 μW for the 532 nm laser and 0.07 μW for the 473 nm laser, respectively. The PL mapping was captured by a piezoelectric driven x–y stage that allowed a positioning precision down to 200 nm (PIMikroMove). For absorption measurement, the sample was transferred onto the quartz substrate and excited by the stabilized tungsten-halogen light source (Thorlabs SLS201L/M). For PLE measurement, the sample was excited with a supercontinuum white light laser (NKT) together with the acoustooptic tunable filter (OYSL photonics). For time-resolved PL measurement, the sample was also excited by the supercontinuum laser with a repetition frequency of 80 MHz (NKT), then synchronized to a single photon counting module (PicoQuant TimeHarp 260) and a single photon detector (Micro Photon Devices). For gate-dependent PL measurement, the gate voltage was applied by the sourcemeter (Keithley 2400). For polarization-resolved PL measurement, a set of quarter-wave plates (Thorlabs SAQWP05M-700 and AQWP05M-580) and a polarizer (Thorlabs WP25M-UB) were used to identify the polarization degree.

**Data availability**

The data supporting the findings of this study are available within the paper and Supplementary Information. Extra data are also available from the corresponding authors upon reasonable request.

**Acknowledgments**

D. L. acknowledges the support from NSFC (61674060), the National Key Research and Development Program of China (2018YFA0704403), and the Innovation Fund of WNLO

**Author contributions**

Y. C. and D. L. designed and convinced the study. Y. C. performed the experiments under the supervision of D. L. Z. L prepared the substrate with electrodes. J. H synthesized 2D perovskites. Y. C. fabricated the samples and carried out the optical measurements. W. Y. and J. L. assisted in the PL measurements. Y. C. and D. L. wrote the manuscript. All authors discussed the results and contributed to the manuscript.

**Competing interests**

The authors declare no competing financial interests.

Figures

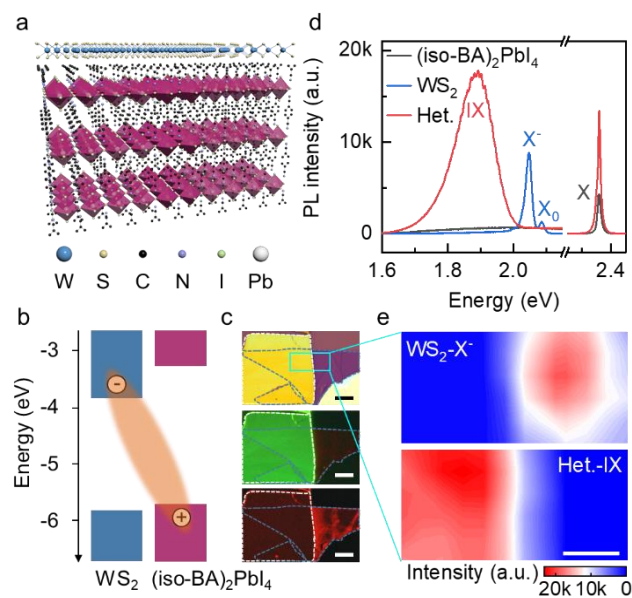

**Fig. 1 IXs in the WS$_2$/(iso-BA)$_2$PbI$_4$ heterostructure. a,** Schematic of a WS$_2$/(iso-BA)$_2$PbI$_4$ heterostructure with the WS$_2$ monolayer on the top of the (iso-BA)$_2$PbI$_4$ flake. **b,** Energy band diagram of monolayer WS$_2$ and (iso-BA)$_2$PbI$_4$. The orange oval shows the interlayer transition from the conduction band of (iso-BA)$_2$PbI$_4$ to the valance band of WS$_2$. **c,** Optical image (top) and fluorescence images under a 520 nm long-pass filter (middle) and a 590 nm long-pass filter (bottom) of the heterostructure at room temperature. The blue and white dashed lines delimit the edges of the WS$_2$ monolayer and (iso-BA)$_2$PbI$_4$ flakes. Scale bars, 10 μm. **d,** PL spectra of the individual (iso-BA)$_2$PbI$_4$, individual WS$_2$, and the heterostructure (Het.) under a 532 nm laser excitation at 78 K. The labels on PL curves refer to excitonic transitions. **e,** PL peak intensity map of the X$^-$ emission of bare WS$_2$ (spectral range 2.03-2.05 eV) and the IX emission of heterostructure (spectral range 1.85-1.89 eV) in the cyan box of **(c)** under a 532 nm laser excitation at 78 K. Scale bar, 10 μm

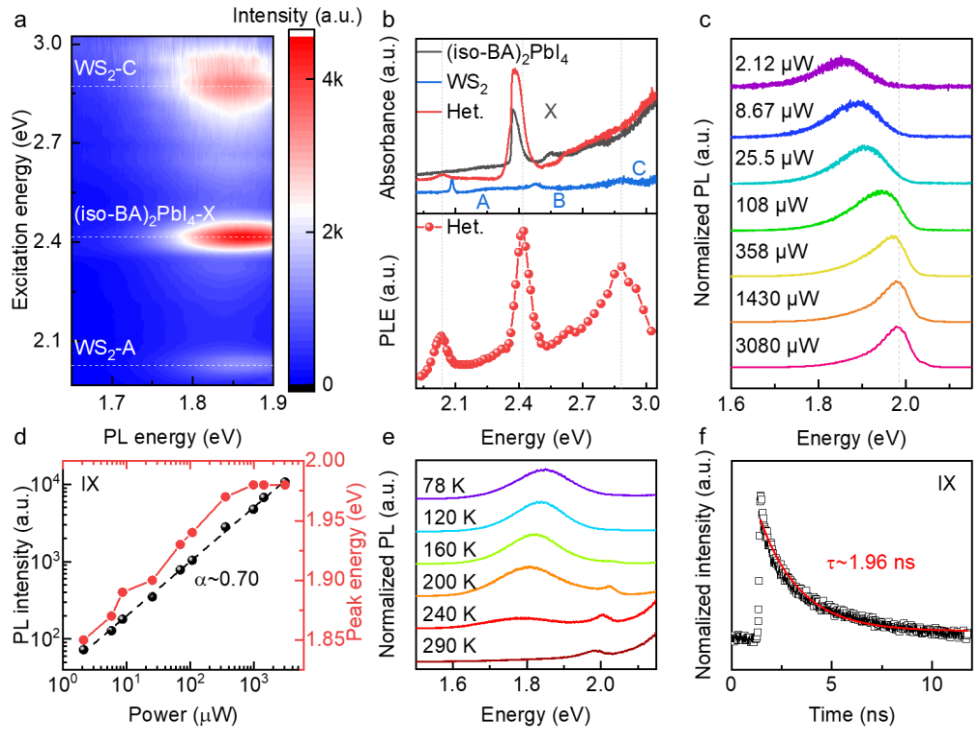

**Fig. 2 Optical characterizations of IXs. a,** PLE intensity map of the WS$_2$/(iso-BA)$_2$PbI$_4$ heterostructure. **b,** Absorption spectra of the heterostructure and individual constituent layers on a quartz substrate, and the PLE spectrum for heterostructure at IX emission energy at 1.86 eV. The vertical light gray lines indicate relevant excitonic transitions. **c,** Power-dependent PL spectra of normalized IXs under a 532 nm laser excitation. The vertical light gray line serves as a guide to the eye for the peak energy shift. **d,** Logarithmic plot of peak energy and intensity of IXs as a function of incident laser power. The fitted black dashed line gives out a sublinear slope of 0.7. **e,** Temperature-dependent PL spectra of normalized IXs under a 532 nm laser excitation. **f,** Time-resolved PL spectrum of IX emission shows a lifetime of about 1.96 ns at 78 K.

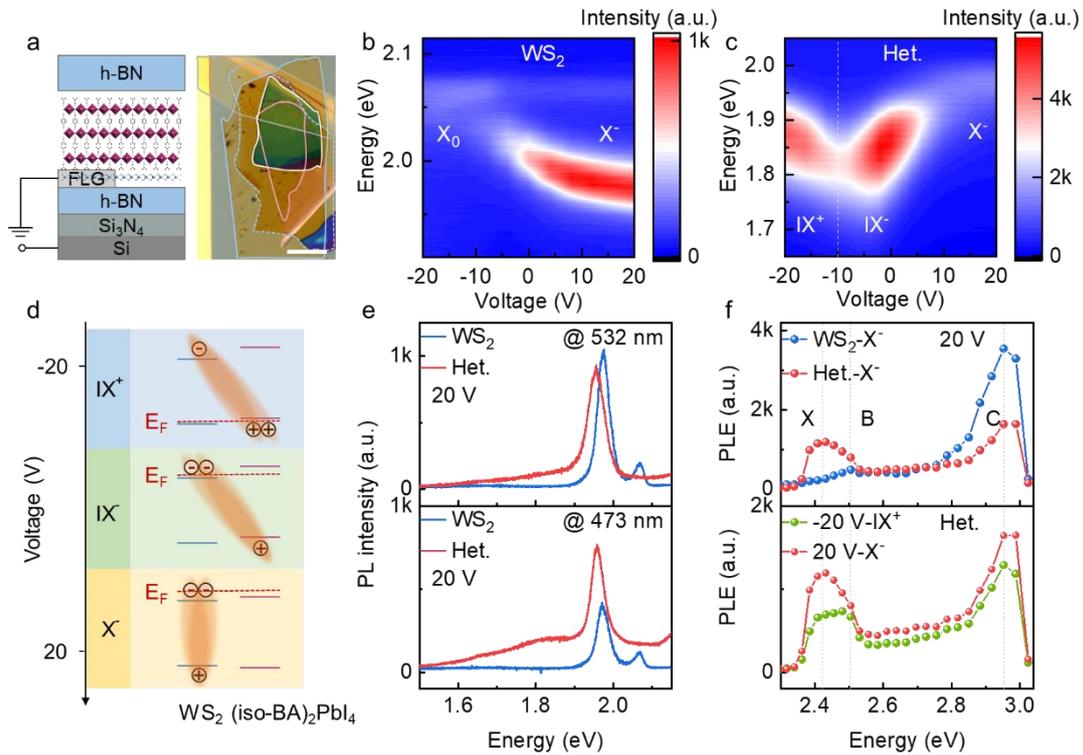

**Fig. 3 Electrostatic doping of IXs. a,** Schematic and optical image of the device structure when WS₂ is directly contacted with FLG. The blue dashed, gray, pink, white, and blue lines delimit the edges of the bottom *h*-BN, FLG, WS₂, (iso-BA)₂PbI₄, and top *h*-BN layers from bottom to top. Scale bar, 10 μm. **b,c,** Color map of PL spectra of individual WS₂ **(b)** and the heterostructure **(c)** as a function of $V_g$ under a 532 nm laser excitation. **d,** Schematic diagram of relative band alignment between WS₂ and (iso-BA)₂PbI₄ with electrostatic doping, along with the Fermi energy level (red dashed line). The orange ovals outline the transitions of different exciton states. **e,** PL spectra of individual WS₂ and the heterostructure for $V_g$= 20 V under the 532 nm (top panel) and the 473 nm (bottom panel) laser excitation. **f,** PLE spectra for WS₂ and heterostructure at X⁻ emission at 1.98 eV for $V_g$= 20 V (top panel), respectively. Comparison of PLE spectra for heterostructure at IX⁺ emission at 1.86 eV for $V_g$= −20 V and at X⁻ emission at 1.98 eV for $V_g$= 20 V (bottom panel).

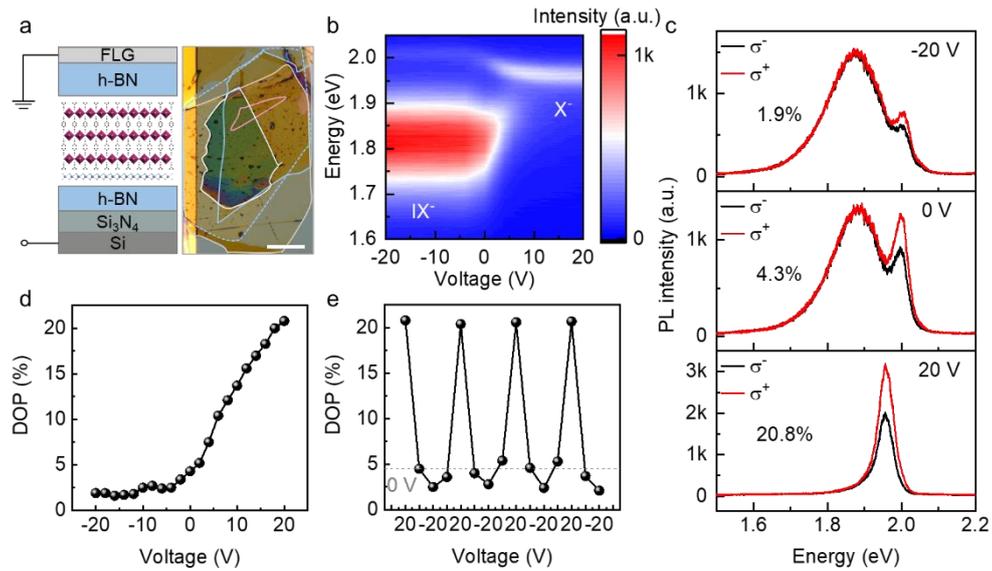

**Fig. 4. Electrically controllable large valley polarization switch mediated by CIXs. a.** Schematic and optical image of the device structure when FLG is covered on the top of *h*-BN encapsulated heterostructure without touching WS$_2$. The blue dashed, pink, white, blue, and gray lines delimit the edges of the bottom *h*-BN, WS$_2$, (iso-BA)$_2$PbI$_4$, top *h*-BN, and FLG layers from bottom to top. Scale bar, 10 μm. **b,** Color map of PL spectra of the heterostructure as a function of V$_g$ under a 532 nm laser excitation. **c,** Polarization-resolved PL spectra of the heterostructure for V$_g$= −20 V. 0 V and 20 V under a σ$^+$ polarized 532 nm laser excitation. **d,** DOP of the heterostructure as a function V$_g$. **e,** Valley polarization switching based on the DOP variation of the heterostructure against the repeated V$_g$ cycles. The gray horizontal dashed line marks the DOP value around 0 V.